\documentclass[12pt]{article}
\usepackage{amssymb,slashed,latexsym}
\newcommand{\ft}[2]{{\textstyle\frac{#1}{#2}}}
\def\Re{\mathop{\rm Re}\nolimits}
\def\Im{\mathop{\rm Im}\nolimits}

\def\rme{{\rm e}}
\def\rmi{{\rm i}}
\def\rmd{{\rm d}}
\newsavebox{\uuunit}
\sbox{\uuunit}
    {\setlength{\unitlength}{0.825em}
     \begin{picture}(0.6,0.7)
        \thinlines
        \put(0,0){\line(1,0){0.5}}
        \put(0.15,0){\line(0,1){0.7}}
        \put(0.35,0){\line(0,1){0.8}}
       \multiput(0.3,0.8)(-0.04,-0.02){12}{\rule{0.5pt}{0.5pt}}
     \end {picture}}

\csname @addtoreset\endcsname{equation}{section}
\newcommand{\SU}{\mathop{\rm SU}}
\newcommand{\SO}{\mathop{\rm SO}}
\newcommand{\U}{\mathop{\rm {}U}}
\newcommand{\USp}{\mathop{\rm {}USp}}
\newcommand{\OSp}{\mathop{\rm {}OSp}}

\newcommand{\Sl}{\mathop{\rm {}S}\ell }

\newif\ifpdf
\ifx\pdfoutput\undefined
   \pdffalse
   \usepackage{cite}
 \else
   \pdfoutput=1
   \pdftrue
  \usepackage[pdftex]{hyperref}
\fi

\begin{document}

\begin{titlepage}
\begin{flushright}
KUL-TF-06/22\\
hep-th/0609048
\end{flushright}
\vspace{.5cm}
\begin{center}
\baselineskip=16pt {\LARGE    Effective supergravity descriptions \\
\vskip 0.2cm of superstring cosmology
}\\
\vfill
{\large  Antoine Van Proeyen, 
  } \\
\vfill
{\small Instituut voor Theoretische Fysica, Katholieke Universiteit Leuven,\\
       Celestijnenlaan 200D B-3001 Leuven, Belgium.
      \\[2mm] }
\end{center}
\vfill
\begin{center}
{\bf Abstract}
\end{center}
{\small
 This text is a review of aspects of supergravity theories that are
 relevant in superstring cosmology. In particular, it considers the
 possibilities and restrictions for `uplifting terms', i.e. methods to
 produce de Sitter vacua. We concentrate on $N=1$ and $N=2$
 supergravities, and the tools of superconformal methods, which clarify
 the structure of these theories. Cosmic strings and embeddings of target
 manifolds of supergravity
 theories in others are discussed in short at the end.}

 \vfill \hrule width 3.cm\vspace{2mm}

{\footnotesize \noindent To be published in the proceedings of the 2nd
international conference on Quantum Theories and Renormalization Group in
Gravity and Cosmology, Barcelona, July 11-15, 2006, Journal of Physics A. \\
e-mail: antoine.vanproeyen@fys.kuleuven.be }
\end{titlepage}
\addtocounter{page}{1}
 \tableofcontents{}
\newpage
\section{Introduction}
 Many people in string theory nowadays study the landscape of vacua that
 are possible in the context of the different string theories and related
 to different compactifications of the 10-dimensional spacetime. The
 resulting vacua each lead to an effective supergravity describing the
low-energy physics close to this vacuum. In this sense, the landscape of
string theories is a landscape of supergravity theories.

The basic superstring theories have already a supergravity as field
theory approximation. After the choice of a compact manifold one is left
with an effective lower dimensional supergravity whose number of
supersymmetries is determined by the Killing spinors of the compact
manifold. Fluxes on branes involved in the string setup and
non-perturbative effects lead to `gauged supergravities'. This are
supergravity theories that are not only the gauge theory of
supersymmetry, but also of an extra ordinary Lie group. The combination
of supersymmetry with these ordinary gauge symmetries requires some
special care.

Though we have discussed here supergravity in the context of superstring
theories, one must warn the reader immediately that not every
supergravity theory can (so far as we know nowadays) be obtained from a
superstring setup. The set of supergravities that is related to
superstring theories has been enlarged in different ways in the past.
E.g. there was a time when one did not know how any gauged supergravity
would be obtained from superstring theory. It is an open question whether
in the future every supergravity theory can be given an embedding in a
superstring theory, but until now it does not look like that will be the
case.

In section \ref{ss:CosmoUplift} we will repeat the problems with
producing de Sitter vacua in supergravity and the related issue of
uplifting terms. Section \ref{ss:supergravities} will give an overview of
the various supergravities. But we will then further concentrate on $N=1$
(and sometimes $N=2$) and give the general structure of these theories in
section \ref{ss:N12SG}. A way in which the structure of these
supergravities can be understood is the superconformal method, as we
explain in section \ref{ss:superconformal}. The final remarks in section
\ref{ss:finalrem} mention the application of supergravity for
construction of  effective theories of cosmic strings and the issue of
embedding of a smaller supergravity theory in a larger one.

\section{Cosmology and uplifting terms}
\label{ss:CosmoUplift}

Supergravity faced already from the early dates a main problem for its
application in cosmology. Cosmological constants can easily be produced,
but its natural scale would be of the order of the fourth power of the
gravitational constant, which is an order of $10^{120}$ too large, a
`world record of discrepancy between theory and experiment'. Apart from
this problem, whose resolution needs massive parameters of another scale,
another main problem is the sign of the cosmological constant. Indeed, we
have to give up the idea of having a supersymmetric vacuum state. This
statement is by now well known, but let us repeat the argument.

It follows from algebraic considerations. If supersymmetry is preserved,
there should be a superalgebra of isometries in the vacuum state. This
superalgebra should contain the de Sitter algebra if the cosmological
constant is positive. Such superalgebras have been classified, and let us
compare the situation for Anti-de Sitter superalgebras with de Sitter
superalgebras in table~\ref{tbl:dSAdS} \cite{Fre:2002pd}.
\begin{table}[ht]
  \caption{\it Superalgebras with bosonic subalgebra a direct product of (anti) de Sitter
  algebra and R-symmetry.}\label{tbl:dSAdS}
\begin{center}
\begin{tabular}{|c|l|rl|}\hline\hline
{\textbf AdS}& superalgebra&\multicolumn{2}{c|}{R-symmetry}\\ \hline
 $D=4$&$ \OSp(N|4) $ &&$ \SO(N) $\\   \hline
$D=5$&$ \SU(2,2|N)  $ &$N\neq 4:$&$  \SU(N)\times \U(1)$  \\
   &               &$N= 4:$& $  \SU(4)$  \\  \hline
$D=6$& F$^2(4)       $ &&$ \SU(2) $  \\       \hline
$D=7$&$ \OSp(8^*|N) $ &$N $ even:&$  \USp(N)$ \\
\hline \hline
{\textbf dS}  & superalgebra & \multicolumn{2}{c|}{R-symmetry} \\
\hline
$D=4$  & $\OSp(m^*|2,2)$ & $m=2$ & $\SO(1,1)$ \\
  &  & $m=4$ & $\SU(1,1)\times  \SU(2)$ \\
  &  & $m=6$ & $\SU(3,1)$ \\
  &  & $m=8$ & $\SO(6,2)$ \\
\hline
$D=5$  & $\SU^*(4|2n)$ & $n=1$ & $\SO(1,1)\times \SU(2)$ \\
  &  & $n=2$ & $\SO(5,1)$ \\
\hline
$D=6$  & F${}^1(4)$ &  & $\SU(2)$ \\
\hline \hline
\end{tabular}
\end{center}
\end{table}
The de Sitter
superalgebras~\cite{Lukierski:1985it,Lukierski:1984,Pilch:1985aw} have
typically a non-compact R-symmetry subalgebra\footnote{We mention here
the superalgebras that are of Nahm's type~\cite{Nahm:1978tg}, i.e., where
the bosonic subgroup is a direct product of the de Sitter algebra and
another simple group, called R-symmetry. More general de Sitter
superalgebras have been classified
in~\cite{D'Auria:2000ec,Ferrara:2001dn}. But they have not been realized
in concrete models, probably because of the appearance of symmetries that
are not in the de Sitter algebra and do not commute with it.}, which
leads to non-definite signs in the kinetic terms, and hence to ghosts.
Therefore, de Sitter vacua can occur in physical supersymmetric theories
only in a phase where supersymmetry is completely broken. This might even
be welcome in view of the fact that supersymmetry breaking is anyhow
necessary to make contact with reality.

As mentioned in the introduction, supergravity is the field theory
corresponding to superstring theory. For calculations with string
theories it is useful to find an effective supergravity description. If
one wants to describe realistic cosmological models, one needs
`uplifting' terms' to raise the value of the cosmological constant. This
is, e.g., one of the main issues in the KKLT \cite{Kachru:2003aw} models.
Another example is the effective theory of the cosmic string.

The Abrikosov-Nielsen-Olesen string model has a vector field $W_\mu(x)$
and a complex scalar $\phi (x)$, charged under the gauge symmetry of the
vector field (coupling constant $g$). It uses as potential
\begin{equation}\label{VANO}
  V= \ft12D^2\equiv  \ft12g^2\left(\xi -\phi^*\phi\right)^2,
\end{equation}
depending on a constant $\xi $. The configuration is independent of one
of the three spatial directions, and, using polar coordinates $(r,\theta
)$ in the remaining 2-plane, it is of the form
\begin{equation}
  \phi (r,\theta )=|\phi |(r)\rme^{\rmi\theta }, \qquad gW_\mu \rmd x^\mu
  =\alpha (r)\rmd\theta,
  \label{stringconfig}
\end{equation}
where the function $|\phi |(r)$ is zero at $r=0$, and thus $D=g\xi $ at
that point, while it goes to a constant value $\sqrt{\xi }$ at infinity,
leading to a vanishing $D$. The vector is determined by the requirement
that the field strength in the plane directions, $F_{12}$ is equal to
$D$. When this model is considered in the context of supersymmetry, it is
a $1/2$ BPS solution~\cite{Dvali:1997bg,Davis:1997bs}. This model can be
embedded in supergravity \cite{Dvali:2003zh,Binetruy:2004hh}
(see~\cite{Becker:1995sp,Edelstein:1996ba} for 3 spacetime dimensions),
due to a conspiracy of the spin connection and the $R$-symmetry
connection.

The model can then be seen as the final state after the D3-brane --
anti-D3-brane annihilation, which leads to a D1 string. The field $\phi $
is the tachyon field, and the so-called Fayet--Iliopoulos term $\xi $
represents the brane-antibrane energy \cite{Dvali:2003zh}. In general it
leads to a positive term in the potential, which is the uplifting that we
need in this case in an effective supergravity model.

\section{Supergravities}
\label{ss:supergravities}

\begin{table}[htbp]
  \caption{\it Supersymmetry and supergravity theories in dimensions 4 to
  11.  An entry represents the possibility to have supergravity
theories in a specific dimension $D$ with the number of (real)
supersymmetries indicated in the top row.  At the bottom is indicated
whether these theories exist only in supergravity, or also with just
rigid supersymmetry.}
  \label{tbl:mapsusy}
\begin{center}\tabcolsep 5pt
  \begin{tabular}{|c|| *{9}{c|} }
\hline
 $D$ &  \multicolumn{2}{c|}{32} & 24  & 20 & \multicolumn{2}{c|}{16}  & 12 & 8 & 4  \\
\hline
11  &  M &  & &  & \multicolumn{2}{c|}{ }  &  &  &  \\
10  &  IIA & IIB & &  &
 I  &  &  &  &  \\
9   &  \multicolumn{2}{c|}{$N=2$ } &  &&
 $N=1$  &  &  &  &  \\
8   &  \multicolumn{2}{c|}{$N=2$ } &  &&
 $N=1$ &  &  &  &  \\
7  &  \multicolumn{2}{c|}{$N=4$ }  &  &&
  $N=2$ & &  &  &  \\
6   & \multicolumn{2}{c|}{$(2,2)$}&$(2,1)$ & &   $(1,1)$  & $(2,0)$ &  & $(1,0)$   &  \\
5  &  \multicolumn{2}{c|}{$N=8$ }  &$N=6$   & &
  \multicolumn{2}{c|}{$N=4$} &  &  $N=2$  &  \\
4   &  \multicolumn{2}{c|}{$N=8$ }  & $N=6$ & $N=5$ &
   \multicolumn{2}{c|}{$N=4$}   &$N=3$   &$N=2$ &  $N=1$  \\
\hline   & \multicolumn{4}{c|}{SUGRA}  &
 \multicolumn{2}{c|}{SUGRA/SUSY} & SUGRA & \multicolumn{2}{c|}{SUGRA/SUSY}  \\
\hline
\end{tabular}
\end{center}
\end{table}

Supergravity exists in many variants. We restrict here to theories where
the terms in the action are at most quadratic in spacetime derivatives.
Table \ref{tbl:mapsusy} plots the possible supergravities for dimensions
larger or equal to 4. A longer discussion on this table and an extension
thereof is given in \cite{VanProeyen:2003zj}. The theories with 16 or
less real components of the supersymmetry operator allow different
additions of matter multiplets to the supergravity multiplet. Suppose now
that we have selected an entry in this table. Thus we have chosen a
dimension $D$ and a number of supersymmetries ($N$ or $Q$ as you like).
Furthermore, assume that for $Q\leq 16$, one has also specified the
number and type of extra multiplets. E.g. in $D=4$, $N=1$ one already
tells that one wants a theory with 1 vector multiplet and 2 chiral
multiplets. How far is the theory then already fixed? In other words:
what has still to be determined in order to completely specify the
action? The answer is different depending on the range of the number $Q$.
\begin{description}
  \item[$32 \geq  Q > 8$.] In this case, the kinetic terms of all the fields are already fixed.
  The only extra information that one needs is the symmetry group that is gauged by the vector fields
  and its action on the scalars. Once this is known, the full action is
  fixed. In particular, the scalar potential of the theory depends on this
  gauging.
  \item[$Q = 8$.] In this case, apart from the gauging, also the kinetic
  terms can still vary. These kinetic terms are restricted, but can still
  depend on some arbitrary prepotential function. The geometry determined
  by the kinetic terms of the scalars falls in a restricted scheme, which
  is called `special geometry'. Once one has chosen the particular
  special geometry and the gauging, the action (and as such e.g. the scalar
  potential) is fixed.
  \item[$Q = 4$.] Also in this case the gauge group and its action on
  scalars has to be determined. Moreover, some arbitrary functions
  determine the kinetic terms. E.g. for the chiral multiplets there is
  the K{\"a}hler potential, and for the vector multiplets the kinetic terms
  are determined by a holomorphic functions of the complex scalars of the
  chiral multiplets. In this case, the potential depends moreover on a superpotential function
  $W$, and in some cases (Abelian gauge groups) on arbitrary constant
  `Fayet-Iliopoulos' constants as the $\xi $ in (\ref{VANO}). There are
  consistency conditions between the choice of these different
  ingredients. The gauging and the Fayet-Iliopoulos constants should be
  compatible in some way with the choice of kinetic terms and
  superpotential.
\end{description}
To illustrate that the restrictions on the kinetic terms, we present here
table \ref{tbl:geometriesPlus8} of scalar manifolds in theories with more
than 8 real supercharges. The theories are ordered as
  in table~\ref{tbl:mapsusy}. For more than 16 supersymmetries, there is only a
  unique scalar manifold,
  while for 16 and 12 supersymmetries there is a number $n$ indicating
  the number of vector multiplets that are included.

\begin{table}[htbp]
  \caption{{\it Scalar geometries in theories with more than 8
  supersymmetries ($4\leq D\leq  9$).} }\label{tbl:geometriesPlus8}
\begin{center}
  $\begin{array}{|@{\hspace{2pt}}c| 
|@{\hspace{2pt}}c |@{\hspace{2pt}}c| c| *{2}{@{\hspace{2pt}}c|} *{1}{c|}
   }
\hline
 D &32 & 24 &20  & \multicolumn{2}{c|}{16} & 12      \\
\hline
 9  & \frac{\Sl(2)}{\SO(2) \otimes
  {\mathrm O}(1,1)}&&&\frac{\mathrm{O}(1,n)}{\mathrm{O}(n)} \otimes  {\mathrm O}(1,1)  &&  \\[3mm]
8  & \frac{\Sl(3)}{\SU(2)}\otimes \frac{\Sl(2)}{\U(1)} &&& \frac{{\mathrm
O}(2,n)}{\U(1)\times
  {\mathrm O}(n)}\otimes  {\mathrm O}(1,1)&  &      \\[3mm]
7  & \frac{\Sl(5)}{\USp(4)} & && \frac{{\mathrm O}(3,n)}{\USp(2)\times
  {\mathrm O}(n)}\otimes  {\mathrm O}(1,1)& &      \\[3mm]
6  & \frac{\mathrm{O}(5,5)}{\USp(4)\times
\USp(4)}&\frac{\SO(5,1)}{\SO(5)} & &
\frac{\mathrm{O}(4,n)}{\mathrm{O}(n)\times SO(4)}\otimes {\mathrm O}(1,1)
 & \frac{\mathrm{O}(5,n)}{\mathrm{O}(n)\times \USp(4)} &     \\[3mm]
5  & \frac{\mathrm{E}_6}{\USp(8)} & \frac{\SU^*(6)}{\USp(6)} &  &
\multicolumn{2}{c|}{\frac{{\mathrm O}(5,n)}{\USp(4)\times
  {\mathrm O}(n)}\otimes  {\mathrm O}(1,1)}   &      \\[3mm]
4  & \frac{\mathrm{E}_7}{\SU(8)} & \frac{SO^*(12)}{\U(6)} &
\frac{\SU(1,5)}{\U(5)} & \multicolumn{2}{c|}{\frac{\SU(1,1)}{\U(1)}\times
\frac{\SO(6,n)}
{\SU(4)\times \SO(n)}}  & \frac{\SU(3,n)}{\U(3)\times \SU(n)}    \\
\hline
\end{array}$
\end{center}
\end{table}

For 8 supersymmetries, the table is
\begin{equation}
\begin{array}{|c|c|c|}
\hline
 D=6 & D=5 & D=4  \\
\hline &&\\[-3mm]
 \frac{O(1,n)}{O(n)}\times QK  & VSR\times QK  & SK  \times QK  \\
\hline
\end{array}
 \label{tblgeomQ8}
\end{equation}
Here, $VSR$ stands for `very special real' geometry, `SK' for special
K{\"a}hler geometry and `QK' for quaternionic-K{\"a}hler geometry. This thus only
specifies a geometric class of manifolds, but they are not uniquely
determined.

We now look at the other issue: the gauge group. The number of generators
of the gauge group is equal to the number of vector fields.\footnote{One
could say less or equal, but considering that any vector transforms at
least as $\delta W_\mu =\partial _\mu \alpha $, even if the $\alpha $
transformation does not act on anything else, we can say that there is a
$\U(1)$ factor.} This counting includes as well vectors in the
supergravity multiplet and those in vector multiplets. In fact, in
general their kinetic terms are mixed and thus they do not have to be
distinguished. The gauge group is in principle arbitrary, but to have
positive kinetic terms gives restrictions on possible non-compact gauge
groups for any supergravity. The symmetries that they gauge act on the
other fields. In particular, it is (a subgroup of) the isometry group of
the scalar manifold.

For most applications in cosmology the final effective theory has only 4
or 8 real supersymmetries. We will now focus our attention to 4
dimensions, and thus to $N=1$ or $N=2$ theories, as well as to the
geometry that describes the kinetic terms of the scalars.

\section{$N=1$ and $N=2$ supergravities}
\label{ss:N12SG}

The multiplets of particles with spin up to 2 that can be considered in
these theories are given in table \ref{tbl:multipletsN12}.
\begin{table}[bp]
  \caption{\it Multiplets of particles in $N=1$ and $N=2$ in 4 dimensions.}\label{tbl:multipletsN12}
  \begin{tabular}{|cll|l|lll|}
\hline
 \multicolumn{3}{|c|}{$N=1$}
  & \  & \multicolumn{3}{c|}{$N=2$} \\
\hline
 graviton m. & $(2,\ft32)$ &   &   & graviton m. & $(2,\ft32,\ft32,1)$ &   \\
 vector mult. & $(1,\ft12)$ &   &   & vector mult. & $(1,\ft12,\ft12,0,0)$ & special K{\"a}hler  \\
 chiral mult. & $(\ft12,0,0)$ & K{\"a}hler  & &hypermult. & $(\ft12,\ft12,0,0,0,0)$ & quaternionic-K{\"a}hler  \\
\hline
\end{tabular}
\end{table}
The actions of the bosonic sector look like (with $\nabla $ the
gauge-covariant derivative)
\begin{eqnarray}
\left.e^{-1}{\cal L}_{N=1}\right|_{\rm bosonic} &=& \ft12 R -\ft14(\Re
f_{\alpha \beta}) F_{\mu \nu }^\alpha F^{\mu \nu \,\beta }
 +\ft 18 e^{-1}\varepsilon ^{\mu \nu \rho \sigma }(\Im f_{\alpha \beta})
 F_{\mu \nu }^\alpha F_{\rho \sigma }^{\beta } \nonumber
\\&&
  -g_{i \bar \jmath}(\nabla_\mu \phi ^i)(\nabla^\mu {\bar
\phi}^{\bar \jmath}) -{\cal V}_{N=1},
 \nonumber\\
 \left. e^{-1}{\cal L}_{N=2}\right|_{\rm bosonic}&=&\ft12 R
+\ft14(\Im {\cal N}_{IJ}) F_{\mu \nu }^I F^{\mu \nu \,J }+\ft 18
e^{-1}\varepsilon ^{\mu \nu \rho \sigma }(\Re {\cal N}_{IJ})
 F_{\mu \nu }^I F_{\rho \sigma }^J
\nonumber\\
 &&  -g_{\alpha \bar \beta }(\nabla_\mu z^\alpha )(\nabla ^\mu \bar z^{\bar \beta
 }) -\ft12 g_{XY}\nabla _\mu q^X\nabla ^\mu q^Y -{\cal V}_{N=2}.
 \label{actionsN12bos}
\end{eqnarray}
For $N=1$, $\alpha $ labels the vector multiplets, and $i,\bar \imath$
label the complex scalars. For $N=2$ the complex scalars of the vector
are labelled by $\alpha ,\bar \alpha=1,\ldots n_V$. The $n_V$ vectors
combine with the vector of the gravity multiplet, leading to the label
$I=0,1,\ldots ,N_V$. The $4n_H$ real scalars of the hypermultiplets are
labelled by the index $X$. In (\ref{actionsN12bos}), $g_{i \bar \jmath}$,
$g_{\alpha \bar \beta }$ and $g_{XY}$ are thus the metrics of the scalar
manifolds mentioned in table \ref{tbl:multipletsN12}. $f_{\alpha \beta }$
are holomorphic functions of the scalars $\phi ^i$, which should just
satisfy some consistency conditions with the gauge group. But the
analogous functions ${\cal N}_{IJ}$ are complex functions that are
determined already by the special K{\"a}hler geometry. This brings us to the
last terms: the scalar potentials.

For $N=1$ the potential is determined by a holomorphic superpotential
$W(\phi )$ and by the gauging. The latter means that the action of the
gauge group on the scalar manifold determines part of the potential. It
was mentioned above that for $N\geq 2$ the potential is only determined
by gauging. This means here that it can be written as a function of the
way in which the gauge group acts on hypermultiplets and vector multiplet
scalars, encoded in the `Killing vectors'. Vector multiplet scalars are
by the structure of the multiplet in the adjoint of the gauge group. That
determines already their Killing vectors. For the hypermultiplets, the
manifold can have a group of isometries, of which a subgroup can be
gauged by the vectors of the vector multiplets. It is a general fact
(Ward identity) in supersymmetry that the scalar potential can be written
as a sum of the square of the scalar part of the supersymmetry
transformations of the fermions, where the way in which the square must
be taken is determined by the metric
\cite{Cecotti:1985mx,Ferrara:1985gj,Cecotti:1986wn}. We illustrate this
here for $N=1$ supergravity. The potential can be written as
\begin{equation}
  V= -3M_P^{-2}F_0\bar F_0+
F_ig^{i\bar\jmath}\bar F_{\bar\jmath}
 +\ft12
D^\alpha(\Re f_{\alpha \beta}) D^\beta,
 \label{VN1d4}
\end{equation}
where $M_P$ is the Planck mass, $F_0$ appears in the supersymmetry
transformation of the gravitino, $F_i$ in that of the chiral fermions and
$D^\alpha $ in that of the gaugini as in
\begin{eqnarray}
 \delta \psi _{\mu L}  & = & \left( \partial _\mu  +\ft14
\omega _\mu {}^{ab}(e)\gamma _{ab} +\ft 12\rmi A_\mu^B \right)\epsilon_L
+\ft12 M_P^{-2}\gamma _\mu F_0
\epsilon _R, \nonumber\\
\delta \chi _i& = & \ft12\slashed{\nabla }\phi_i \epsilon _R- \ft12
F_i\epsilon _L, \qquad \delta \lambda^\alpha  =\ft14\gamma ^{\mu \nu }
F_{\mu \nu }^\alpha\epsilon +\ft12\rmi \gamma _5 D^\alpha \epsilon .
 \label{susyfermionsN1d4}
\end{eqnarray}
The quantities $A_\mu ^B$, $\omega _\mu {}^{ab}(e)$, and the
$\hat{\partial }$ depends on fields in a way that is not relevant here
(see e.g. \cite{Binetruy:2004hh}). The first two terms in (\ref{VN1d4})
are called the $F$-terms and the last term is the $D$-term. The former
depend on the superpotential $W$ (and on the K{\"a}hler potential at order
$M_P^{-2}$). The $D$-term depends on the gauge transformations of the
scalars (and also on the K{\"a}hler potential) and can depend on arbitrary
Fayet-Iliopoulos constants $\xi _\alpha $ for $\U(1)$ factors.

\section{Superconformal methods}
\label{ss:superconformal}

The superconformal method offers a simplification of supergravity by
using a parent rigid supersymmetric theory. A conceptual difference
between supersymmetry and supergravity is that the concept of multiplets
is clear in supersymmetry but they become mixed in supergravity. This
makes superfields an easy tool for rigid supersymmetric theories, but
much more complicated for supergravity. The main idea of the
superconformal methods is that the supergravity theory can be obtained
starting by a rigid supersymmetric theory that has conformal symmetry.
This conformal symmetry becomes part of a superconformal group that is
gauged. Then, the extra conformal symmetries (and their superpartner
symmetries) are gauge-fixed. Before the gauge fixing, everything looks
like in rigid supersymmetry with covariantizations. After the gauge
fixing it becomes a Poincar{\'e} supergravity theory.

This is the generalization of starting with the Lagrangian
\begin{equation}
{\cal L}= -\ft12\sqrt{g}\phi \Box^C\phi=-\ft12\sqrt{g}\phi \Box\phi
+\ft1{12}\sqrt{g}R\phi ^2.
 \label{Lconfscalar}
\end{equation}
The conformal covariant D'Alembertian contains the well known $R\phi ^2$
term. This starting action is invariant under local conformal
transformations, which scale the metric and also the scalar as $\delta
\phi(x) = \Lambda _D(x)\phi (x)$. Hence one can fix the value of this
scalar as gauge fixing of these dilatations. A convenient value is $\phi
=\sqrt{6}M_P$, which reduces the Lagrangian (\ref{Lconfscalar}) to the
Einstein-Hilbert form:
\begin{equation}
  {\cal L}=\frac{M_P^2}{2}\sqrt{g}R.
 \label{LagrEH}
\end{equation}
The gauge-fixed value of the scalar thus determines the Planck scale.
Remark that to start this example we started with the action of the
scalar with a sign such that the kinetic terms are negative definite.
This leads to positive kinetic energy of the graviton. In general, when
there are more scalars, one has to start with a theory with signature of
the kinetic energies $(-+\ldots +)$, and the gauge fixing procedure will
remove the negative signature scalar.

To generalize this to supersymmetry, we have to extend the conformal
group of transformations to a superconformal group. This enlarges also
the set of bosonic symmetries. Let us concentrate here on $N=1$ in 4
dimensions. In that case this superconformal group includes apart from
the conformal group (including the dilatations as above) also a $\U(1)$
R-symmetry. The name R-symmetry refers to the fact that this symmetry
does not commute with supersymmetry. To obtain Poincar{\'e} supergravity one
has then to start with a Weyl multiplet and a compensating chiral
multiplet. The Weyl multiplet contains the set of gauge fields of the
superconformal algebra. It is the generalization of the $g_{\mu \nu }$
field above. The chiral multiplet is the generalization of the field
$\phi $ in the above example. In this chiral multiplet sits a complex
field, which we denote by $Y$, whose modulus is now fixed by gauge fixing
of dilatations and its phase is fixed by the gauge fixing of the
mentioned $\U(1)$.

When one wants to describe the coupling of $n$ chiral multiplets to
supergravity, one starts with rigid supersymmetry where $n+1$ chiral
multiplets appear. For rigid supersymmetry, the kinetic terms of the
scalars should define a K{\"a}hler manifold. Now we impose that there is a
conformal symmetry. Technically, this is phrased as the presence of a
`closed homothetic Killing vector' $k$. Then the structure of the K{\"a}hler
manifold implies automatically another Killing vector $Jk$, where $J$ is
the complex structure of the manifold. This is gauged by the gauge field
of the $\U(1)$ in the Weyl multiplet. The gauge fixing of the dilatations
and of the $\U(1)$ then lead to a $n$-dimensional Hodge-K{\"a}hler manifold,
which is the geometric structure of chiral multiplets coupled to
supergravity. More details can be found e.g. in
\cite{VanProeyen:1983wk,Kallosh:2000ve,Binetruy:2004hh}.

As an example of the simplifications and the way in which the conformal
setup clarifies the structure of the theory, let us consider the scalar
potential. The $F$-term 
is
\begin{equation}
   V_F={\rm e}^{(K/M_P^2)}\left[ -3M_P^{-2}W \bar W + (D_i W) g^{i\bar j
}(D_{\bar j }\bar W)\right],\qquad
 D_i W \equiv  \partial _i W +M_P^{-2}(\partial _i K) W,
 \label{VFterm}
\end{equation}
where $K$ is the K{\"a}hler potential and $W$ the superpotential in the
supergravity formulation. In the superconformal setup this is unified by
denoting all the scalars as $Z^A$, which thus includes the compensator
$Y$ and the other scalars $z^i$. Then it is
\begin{eqnarray}
 V_F&=&  \left.(\partial _A{\cal W}) G^{A\bar B} (\partial _{\bar B}{\cal
  W})\right|_{{\cal K}=-3M_P^2}\nonumber\\
&&{\cal K}= -3Y\bar Y {\rm e}^{-K(z,\bar z)/(3M_P^2)},\qquad G_{A\bar
B}=\partial _A\partial _{\bar B}{\cal K},\qquad  {\cal W}= Y^3M_P^{-3}
W(z).
 \label{VFconf}
\end{eqnarray}
${\cal K}$ is the K{\"a}hler potential and ${\cal W}$ is the superpotential
of the rigid theory, and the gauge fixing of dilatations has fixed the
value of ${\cal K}$ to $-3M_P^2$. A similar simplification occurs for the
value of the $D$-term in (\ref{VN1d4}) (assuming here a gauge-invariant
K{\"a}hler potential)
\begin{eqnarray}
  D^\alpha &=&(\Re f_{\alpha \beta })^{-1}{\cal P}_\beta ,\qquad
  \nonumber\\
  {\cal P}_\alpha&=& \ft12\rmi k^i_\alpha\partial_iK
-\ft12\rmi k^{\bar i}_\alpha\partial_{\bar i} K
   +g\xi_\alpha =\left(\ft12\rmi k^A_\alpha\partial_A
{\cal K} -\ft12\rmi k^{\bar A}_\alpha\partial_{\bar A} K \right)_{{\cal
K}=-3M_P^2}.
 \label{Dfromconf}
\end{eqnarray}
In the first expression for ${\cal P}$ occur the Killing vectors in the
directions of the physical scalars $k_\alpha^i$ and the Fayet-Iliopoulos
term $\xi_\alpha $. The latter is in the conformal formulation related to
the component of the Killing vector in the direction of $Y$, i.e. $
k_\alpha^Y= \rmi g \xi_\alpha Y/(3M_P^2)$.

\section{Final remarks}
\label{ss:finalrem}

We have given an overview of a landscape of possible supergravity
theories. Consistency with supergravity gives also many restrictions on
an effective field theory of the fields near a string theory vacuum. We
have illustrated here some general features of supergravity theories, and
given the key ingredients of the superconformal formulation that leads to
insights in their structure.

This has been used to construct the effective $N=1$ supergravity theory
of cosmic strings in \cite{Dvali:2003zh}. The fact that this construction
is embedded in the basic $N=1$ theory may look as a shortcoming. However,
often such models can be embedded in larger supergravity theories. E.g.
the mentioned cosmic string configuration could be embedded in an $N=2$
supergravity \cite{Achucarro:2005vz} in a way such that effectively only
some fields of the larger theory are non-vanishing, which brings us back
to the $N=1$ setup. Such embeddings are, however, nontrivial. There is a
main consistency requirement that the reduction of the field equations of
the $N=2$ theory to the subsector gives the same result as calculating
the field equations of the reduced Lagrangian. $N=2$ consistent
truncations have been considered in detail in e.g.
\cite{Andrianopoli:2001zh,Andrianopoli:2001gm}. In geometrical terms it
says that the submanifold must be `geodesic'. This means that any
geodesic in the submanifold should be a geodesic of the ambient manifold.

Such principles also hold when one considers a small set of multiplets in
a supergravity and wants to consider it within a larger model with a
larger set of multiplets. E.g. recently we investigated how several $N=2$
supergravities with special geometry can be embedded in each other, such
that a solution of the smallest one can be taken over as a solution of
the larger one. That lead to a restricted set of basic homogeneous
special geometries \cite{Fre:2006eu}.

\medskip
\section*{Acknowledgments.}

\noindent I thank the organizers of the IRGAC workshop for the nice and
stimulating environment of this meeting. This work is supported in part
by the European Community's Human Potential Programme under contract
MRTN-CT-2004-005104 `Constituents, fundamental forces and symmetries of
the universe', and in part by the FWO - Vlaanderen, project G.0235.05 and
by the Federal Office for Scientific, Technical and Cultural Affairs
through the "Interuniversity Attraction Poles Programme -- Belgian
Science Policy" P5/27.

\providecommand{\href}[2]{#2}\begingroup\raggedright\endgroup

\end{document}